\documentclass[12pt]{article}
\usepackage{graphicx}
\usepackage{amssymb}

\title{Orbital magnetic moment of the electron in the hydrogen atom in deformed space with minimal length}
\author{M.M. Stetsko\footnote{E-mail: mykola@ktf.franko.lviv.ua}\ \ and
V.M. Tkachuk\footnote{E-mail: tkachuk@ktf.franko.lviv.ua}
\\
  {\small Department of Theoretical Physics, Ivan Franko National University of Lviv,}\\
{\small 12 Drahomanov St., Lviv, UA-79005, Ukraine}}
\begin{document}
\maketitle

\abstract{We investigated the orbital magnetic moment of electron
in the hydrogen atom in deformed space with minimal length. It
turned out that corrections to the magnetic moment caused by
deformation depend on one parameter in the presence of
two-parametric deformation. It is interesting to note that the
correction to orbital magnetic moment is similar to the correction
that follows from relativistic theory but it has an opposite sign.
Using the upper bound for minimal length obtained in previous
papers we estimated the upper bound for relative correction to
orbital magnetic moment and obtained the value $\sim 10^{-12}$.
This is four power less than the relative error for most recent
experimental values of Bohr magneton.}

\section{Introduction}

In recent years a lot of attention has been devoted to quantum
mechanics with a deformed commutation relations. This interest was
impelled by several independent lines of investigation such as
string theory and quantum gravity which suggested the existence of
a finite lower bound to the possible resolution of length (minimal
length) \cite{gross,maggiore,witten}. Kempf \textit{et al.} showed
that minimal length could be obtained as a minimal uncertainty in
position from the deformed commutation relations
\cite{kempf1,kempf2,kempf3,kempf4,kempf5}. In \cite{hossenfelder}
it was shown that generalized commutation relations leading to the
existence of minimal length could be obtained from modified
dispersion relations. We also note that for the first time the
deformed algebra leading to a quantized space-time was introduced
by Snyder in the relativistic case \cite{snyder}. In the
$D$-dimensional case deformed algebra proposed by Kempf takes the
form:
\begin{eqnarray}\label{algebra}
\begin{array}{l}
[X_i, P_j]=i\hbar(\delta_{ij}(1+\beta P^2)+\beta'P_iP_j), [P_i,
P_j]=0,
\\
\\
\displaystyle {[X_i,
X_j]}=i\hbar\frac{(2\beta-\beta')+(2\beta+\beta')\beta
P^2}{1+\beta P^2}(P_iX_j-P_jX_i),
\end{array}
\end{eqnarray}
where $\beta$ and $\beta'$ are the parameters of deformation. We
also suppose that parameters of deformation are nonnegative
$\beta,\beta'\geqslant 0$. Having the uncertainty relation one can
obtain that minimal length equals $\hbar\sqrt{\beta+\beta'}$. We
note that in the special case $2\beta=\beta'$ the position
operators in linear approximation over the deformation parameters
commute, i.e. $[X_i,X_j]=0$.

Deformed commutation relations (\ref{algebra}) bring new
difficulties in quantum mechanics. Only a few problems are known
for which the energy spectra have been found exactly. They are
one-dimensional harmonic oscillator with minimal uncertainty in
position \cite{kempf2} and also with minimal uncertainty in
position and momentum \cite{Tkachuk1,Tkachuk2}, $D$-dimensional
isotropic harmonic oscillator \cite{chang, Dadic},
three-dimensional relativistic Dirac oscillator \cite{quesne} and
one-dimensional Coulomb problem \cite{fityo}.

The hydrogen atom is the key one in modern physics. Hydrogen atom
allows not only highly accurate theoretical predictions but it is
also well studied experimentally offering the most precisely
measured quantities. The hydrogen atom problem in deformed space
with minimal length was considered for the first time by Brau in
the special case $2\beta=\beta'$ \cite{Brau}. The general case of
deformation $2\beta\neq\beta'$ was investigated in \cite{Benczik}.
The authors used perturbation theory and calculated corrections to
the energy levels. But the perturbation theory proposed by the
authors did not allow to obtain corrections to the $s$-levels. To
avoid this problem the authors used the numerical methods and
cut-off procedure. In our work \cite{mykola} we developed the
modified perturbation theory enabling to calculate the corrections
for arbitrary energy levels in hydrogen atom including $s$-levels.
In \cite{mykola2} we applied the modified theory for finding the
corrections to the $ns$-levels in the hydrogen atom. The hydrogen
atom in a deformed space was also considered in
\cite{hossenfelder}. In work \cite{mykola3} was considered the
elastic scattering problem on the Yukawa and Coulomb potential in
deformed space with minimal length.

In this work we proceed the examination of hydrogen atom in
deformed space with minimal length. Using the results obtained in
works \cite{mykola,mykola3} we investigate the orbital magnetic
moment of the electron. This paper is organized as follows. In the
second section we obtain the continuity equation for the particle
in the Coulomb field taking into account some of the results of
work \cite{mykola3}. In the third section we consider the
corrections to the wave function of the hydrogen atom in a
deformed space. In the fourth section we calculate the orbital
magnetic moment and compare our results with relativistic
corrections. And finally the fifth section contains the
discussion.

\section{Continuity equation}
Here we consider the motion of electron in the Coulomb field of
the nucleus. In work \cite{mykola3} we found the continuity
equation in a more general case for the particle in Yukawa field.
Using the results of work \cite{mykola3} we can immediately write
the continuity equation for the Coulomb field. But for the Coulomb
potential the calculations are considerably simpler than for
Yukawa field and we give these calculations here.

The Hamiltonian for a particle in the external Coulomb field reads
\begin{equation}\label{hamiltonian}
H=\frac{\textbf{P}^2}{2M}-\frac{e^2}{R},
\end{equation}
where operators of position $X_i$ and momentum $P_i$ satisfy the
deformed commutation relations (\ref{algebra}) and
$R=\sqrt{\sum^3_{i=1}X^2_i}$

To construct the continuity equation we write the Schr\"{o}dinger
equation
\begin{equation}\label{Schroedinger}
i\hbar\frac{\partial\psi}{\partial t}=H\psi
\end{equation}

One can write the following relation using equation
(\ref{Schroedinger})
\begin{equation}\label{cont1}
\frac{\partial\rho}{\partial t}=\frac{1}{i\hbar}(\psi^*H\psi-\psi
H\psi^*),
\end{equation}
where $\rho=|\psi|^2$.

To construct the continuity equation it is necessary to use the
representation of the operators of positions and momenta that
satisfy the deformed commutation relations (\ref{algebra}). The
momentum representation for such an algebra is well known, but it
is not convenient for us. We use the following representation that
obeys algebra (\ref{algebra}) in the first order over $\beta$,
$\beta'$
\begin{eqnarray}\label{rep1}
\left\{
\begin{array}{l}
 X_i=x_i+\frac{2\beta-\beta'}{4}\left(x_ip^2+p^2x_i\right),
\\
P_i=p_i+\frac{\beta'}{2}p_ip^2;
\end{array}
\right.
\end{eqnarray}
where $p^2=\sum^3_{j=1}p^2_j$ and operators $x_i$, $p_j$ satisfy a
canonical commutation relation. The position representation
$x_i=x_i$, $p_j=i\hbar\frac{\partial}{\partial x_j}$ can be taken
for the ordinary Heisenberg algebra.

As was shown in work \cite{mykola} Hamiltonian (\ref{hamiltonian})
can be expressed in the following form using representation
(\ref{rep1}) and taking into account only the first order terms in
$\beta$, $\beta'$
\begin{equation}\label{H3}
H=\frac{p^2}{2M}+\frac{\beta'p^4}{2M}-e^2\left(\frac{1}{\sqrt{r^2+b^2}}-\frac{2\beta-\beta'}{4}
\left(\frac{1}{r}p^2+p^2\frac{1}{r}\right)\right),
\end{equation}
where $b=\hbar\sqrt{2\beta-\beta'}$.

So we can rewrite equation (\ref{cont1}) using the Hamiltonian
(\ref{H3})
\begin{eqnarray}
\frac{\partial\rho}{\partial
t}=\frac{1}{i\hbar}\left(\frac{1}{2M}(\psi^*p^2\psi-\psi
p^2\psi^*)+\frac{\beta'}{2M}(\psi^*p^4\psi-\psi
p^4\psi^*)+\right.\nonumber
\\
\\\nonumber
\left.\frac{(2\beta-\beta')e^2}{4}\left(\psi^*\left(\frac{1}{r}p^2+p^2\frac{1}{r}\right)\psi-
\psi\left(\frac{1}{r}p^2+p^2\frac{1}{r}\right)\psi^*\right)\right),
\end{eqnarray}

Then using explicit form for the operator $p^2=-\hbar^2\nabla^2$
we write the last equation as follows
\begin{eqnarray}\label{cont_equ}
\frac{\partial\rho}{\partial
t}=\frac{1}{i\hbar}\nabla\left(-\frac{\hbar^2}{2M}(\psi^*\nabla\psi-\psi\nabla\psi^*)+
\frac{\beta'\hbar^4}{2M}(\psi^*\nabla^3\psi-\psi\nabla^3\psi^*-\nabla\psi^*\nabla^2\psi+
\right.\nonumber
\\
\\\nonumber
\left.\nabla\psi\nabla^2\psi^*)-\frac{\hbar^2e^2(2\beta-\beta')}{4}\left(\psi^*\left(\frac{1}{r}\nabla+\nabla\frac{1}{r}\right)\psi
-\psi\left(\frac{1}{r}\nabla+\nabla\frac{1}{r}\right)\psi^*\right)\right)
\end{eqnarray}

which can be represented in the continuity equation form:
\begin{equation}
\frac{\partial\rho}{\partial t}+\rm{div}\textbf{j}=0,
\end{equation}
where
\begin{eqnarray}\label{density_flux}
\textbf{j}=\frac{1}{i\hbar}\left(-\frac{\hbar^2}{2M}(\psi^*\nabla\psi-\psi\nabla\psi^*)+
\frac{\beta'\hbar^4}{2M}(\psi^*\nabla^3\psi-\psi\nabla^3\psi^*-\nabla\psi^*\nabla^2\psi+\right.\nonumber
\\
\\\nonumber
\left.\nabla\psi\nabla^2\psi^*)-\frac{\hbar^2e^2(2\beta-\beta')}{4}\left(\psi^*\left(\frac{1}{r}\nabla+\nabla\frac{1}{r}\right)\psi
-\psi\left(\frac{1}{r}\nabla+\nabla\frac{1}{r}\right)\psi^*\right)\right)
\end{eqnarray}
 is the probability density flux and here $\nabla^3=(\nabla,\nabla)\nabla$.

Expression (\ref{density_flux}) for the probability density flux
in a deformed case for the particle moving in the external Coulomb
field is somewhat different from the density flux in the ordinary
quantum mechanics. In contrast to the ordinary quantum mechanics
in a deformed case we have two additional terms into the
continuity equation. One of them is caused by the deformed kinetic
energy. The second contribution is caused by the Coulomb field.
But one should note that in the special case $2\beta=\beta'$ when
the position operators are commutative, i.e. $[X_i,X_j]=0$, the
Coulomb potential does not make any contribution to the continuity
equation.

\section{Corrections to wave function of the hydrogen atom}

When we calculate only the first-order corrections to the energy
spectrum we use the eigenfunctions of undeformed quantum problem
but for high order corrections to the energy spectrum it is
necessary to have the corrections to undeformed wave functions
too. In contrast to this when we calculate the averages for the
other operators but not for the Hamiltonian we must take into
account corrections to the wave functions in the first order of
perturbation. So to obtain the correct expression for the current
density for the electron in the hydrogen atom we must take into
account also corrections to the eigenfunctions caused by
deformation.

The wave function of the perturbed quantum system in the first
order takes the form
\begin{equation}\label{wavefunction}
\psi^{(1)}_{(q)}=\psi^{(0)}_{(q)}+\sum_{(q')\neq(q)}
\frac{V_{(qq')}}{E^{(0)}_{(q)}-E^{(0)}_{(q')}}\psi^{(0)}_{(q')},
\end{equation}
where $\psi^{(0)}_{(q)}$ is the wave function of the unperturbed
system, $E^{(0)}_{(q)}$ is the energy of the unperturbed problem,
$V_{(qq')}$ is the matrix element for the perturbation operator
and $(q)$ is the multiindex.

In our case the unperturbed Hamiltonian is the Hamiltonian of the
ordinary hydrogen atom and the perturbation operator takes the
form similarly to \cite{mykola}
\begin{equation}\label{perturb_oper}
V=\frac{\beta'p^4}{2M}-e^2\left(\frac{1}{\sqrt{r^2+b^2}}-\frac{1}{r}\right.
\left.-\frac{2\beta-\beta'}{4}\left(\frac{1}{r}p^2+p^2\frac{1}{r}\right)\right).
\end{equation}

So using the eigenfunction of an ordinary hydrogen atom we
calculate matrix elements for the perturbation operator
\begin{equation}
V_{(qq')}=\langle
n'l'm'|V|nlm\rangle=\delta_{ll'}\delta_{mm'}\left(2M\beta'{E^{(0)}_n}^2\delta_{n'n}-V_{nn'}\right),
\end{equation}
where $E^{(0)}_n=-\frac{e^2}{an^2}$ is the energy of an ordinary
hydrogen atom and
\begin{eqnarray}\label{matrix_element}
V_{nn'}=\langle n'lm|V|nlm\rangle=-\langle
n'lm|\frac{e^2}{\sqrt{r^2+b^2}}|nlm\rangle+\langle
n'lm|\frac{e^2}{r}|nlm\rangle+\nonumber
\\
\\
\frac{M}{2}(2\beta+3\beta')(E^{(0)}_n+E^{(0)}_{n'})\langle
n'lm|\frac{e^2}{r}|nlm\rangle+M(2\beta+\beta')\langle
n'lm|\frac{e^4}{r^2}|nlm\rangle .\nonumber
\end{eqnarray}
Matrix elements $V_{nn'}$ depend on the orbital quantum number $l$
and do not depend on the magnetic quantum number $m$.

 So the wave function of the hydrogen atom in a deformed space with
minimal length takes the following form
\begin{equation}\label{def_wavefunction}
\psi^{(1)}_{nlm}=\psi_{nlm}+\sum_{n'\neq
n}\frac{V_{nn'}\psi_{n'lm}}{E^{(0)}_n-E^{(0)}_{n'}}\nonumber
\end{equation}
where $V_{nn'}$ is a matrix element given by expression
(\ref{matrix_element}) and
$\psi_{nlm}=R_{nl}(r)P^{|m|}_l(\cos{\vartheta})e^{im\varphi}$ is
the eigenfunction of a undeformed hydrogen atom. We do not give
here the explicit expression for matrix elements because they are
not used in our calculations.

 We would also like to stress that
for the excited states with the nonzero orbital quantum number
$l\neq 0$ we can use a simpler form for the perturbation operator
instead of (\ref{perturb_oper})
\begin{equation}\label{perturb_oper2}
V=\frac{\beta'p^4}{2M}+\frac{(2\beta-\beta')e^2}{4}\left(
\frac{1}{r}p^2+p^2\frac{1}{r}+\frac{2\hbar^2}{r^3}\right).
\end{equation}

As we see this perturbation operator is linear over the
deformation parameters so the correction to the wave function will
be linear in deformation too. We note that for the $s$-states we
are forced to use perturbation operator (\ref{perturb_oper})
because the term proportional to $1/r^3$ in (\ref{perturb_oper2})
gives a divergent contribution in this case. The magnetic moment
has a nonzero value only for excited states. Therefore, for its
calculation we can use just (\ref{perturb_oper2}).

\section{Magnetic moment}

Having a relation for the probability density flux we can find the
magnetic moment of the electron in atom. We calculate the electric
current density for the electron in the atom multiplying
expression (\ref{density_flux}) by the electron charge and using
the wave function of hydrogen atom (\ref{def_wavefunction}).
Taking into consideration only the first order corrections we
obtain
\begin{eqnarray}\label{current_hydrogen}
\textbf{j}_e=-\frac{e}{i\hbar}\left(-\frac{\hbar^2}{2M}({\psi^*_{nlm}}{\bf\nabla}\psi_{nlm}-
\psi_{nlm}{\bf \nabla}\psi^*_{nlm})+
\frac{\beta'\hbar^4}{2M}(\psi^*_{nlm}\nabla^3\psi_{nlm}-\psi_{nlm}\nabla^3\psi^*_{nlm}
-\nabla\psi^*_{nlm}\nabla^2\psi_{nlm}\right.\nonumber
\\\nonumber
\\
\left.+\nabla\psi_{nlm}\nabla^2\psi^*_{nlm})-
\frac{\hbar^2e^2(2\beta-\beta')}{4}\left(\psi_{nlm}^*\left(\frac{1}{r}\nabla+\nabla\frac{1}{r}\right)\psi_{nlm}
-\psi_{nlm}\left(\frac{1}{r}\nabla+\nabla\frac{1}{r}\right)\psi^*_{nlm}\right)\right.
\\\nonumber
\\
\left.-\frac{\hbar^2}{2M}\sum_{n'\neq
n}\frac{V_{nn'}}{E^{(0)}_n-E^{(0)}_{n'}}(\psi^*_{nlm}\nabla\psi_{n'lm}+\psi^*_{n'lm}\nabla\psi_{nlm}-
\psi_{nlm}\nabla\psi^*_{n'lm}-\psi_{n'lm}\nabla\psi^*_{nlm})\right).\nonumber
\end{eqnarray}
We choose the electron charge in the form $-e$, where
$e=4.8203\times 10^{-10}$ is the absolute value of the electron
charge.

It is convenient to calculate the component of the current density
in the spatial spherical coordinates for which the
$\nabla$-operator takes the following form
\begin{equation}\label{nabla}
{\bf{\nabla}}=\textbf{e}_1\frac{\partial}{\partial
r}+\textbf{e}_2\frac{1}{r}\frac{\partial}{\partial\vartheta}+
\textbf{e}_3\frac{1}{r\sin{\vartheta}}\frac{\partial}{\partial\varphi},
\end{equation}
where $\textbf{e}_1$, $\textbf{e}_2$, $\textbf{e}_3$ are the unit
vectors tangent to the coordinate curves at the given point. Using
representation (\ref{nabla}) for the components of the density
current in the spherical coordinates we have
\begin{equation}\label{current_radial}
j_r=0,
\end{equation}
\begin{equation}\label{current_azimuthal}
j_{\vartheta}=0,
\end{equation}
\begin{equation}\label{current_polar}
j_{\varphi}=-\frac{e\hbar
m}{Mr\sin{\vartheta}}\left(|\psi_{nlm}|^2+4\beta'M\psi_{nlm}^*\hat{K}\psi_{nlm}+
2e^2(2\beta-\beta')M\frac{|\psi_{nlm}|^2}{r}+2\sum_{n'\neq
n}\frac{V_{nn'}\psi^*_{n'lm}\psi_{nlm}}{E^{(0)}_n-E^{(0)}_{n'}}\right),
\end{equation}
where $\hat{K}=-\frac{\hbar^2\nabla^2}{2M}$ is the kinetic energy
operator in ordinary quantum mechanics and $m$ is the magnetic
quantum number.

Relations (\ref{current_radial}) and (\ref{current_azimuthal}) can
be obtained immediately when we take into account that functions
$R_{nl}(r)$ and $P^{|m|}_l(\cos{\vartheta})$ are real. This result
coincides with the expression for the components of the current
density vector in ordinary quantum mechanics.

Having a relation for the density current we can calculate the
magnetic moment caused by them
\begin{equation}
\textrm{d}\mu_z=\frac{1}{c}S\ \textrm{d}I=\frac{1}{c}\pi
j_{\varphi}r^2\sin^2{\vartheta}\ \textrm{d}\sigma,
\end{equation}
where $S=\pi r^2\sin^2{\vartheta}$ is the area of the current loop
${\rm d}I=j_{\varphi}{\rm d}\sigma$.

To obtain the total magnetic moment it is necessary to integrate
over all of the current tubes. So we have
\begin{equation}\label{magn_moment}
\mu_z=-\frac{e\hbar m}{2Mc}\left(1+4\beta'M\langle\hat{K}\rangle+
2(2\beta-\beta')e^2M\left\langle\frac{1}{r}\right\rangle \right)
\end{equation}
We note that corrections to the wave function do not make a
contribution to the magnetic moment because in the expression for
the density current (\ref{current_polar}) we have the product of
two orthogonal functions and after integration these terms
disappear.

As was shown in \cite{kempf2} the angular momentum in deformed
space with the minimal length takes the form
\begin{equation}\label{angul_mom_def}
L_{ij}=\frac{1}{1+\beta P^2}(P_iX_j-P_jX_i),
\end{equation}
and operators $X_i$, $P_i$ satisfy algebra (\ref{algebra}).

Substituting representation (\ref{rep1}) in (\ref{angul_mom_def})
and taking into account only the first order terms over the
deformation parameters we obtain
\begin{equation}\label{angul_mom_def}
L_{ij}=p_ix_j-p_jx_i,
\end{equation}
and thus in this case the deformed angular momentum coincides with
an ordinary one.

 We know that for the hydrogen atom the following relations
$\langle\hat{K}\rangle=\langle\hat{p}^2/2M\rangle=e^2/(2an^2)$ and
$\langle e^2/r\rangle=e^2/(an^2)=2\langle\hat{K}\rangle$ take
place therefore we can rewrite expression (\ref{magn_moment}) for
the magnetic moment in the form
\begin{equation}\label{magn_moment2}
\mu_z=-\frac{e}{2Mc}(1+8\beta M\langle \hat{K}\rangle)L_z
\end{equation}

Substituting the explicit form for the averages and taking that
$L_z=\hbar m$ we obtain
\begin{equation}\label{magn_moment3}
\mu_z=-\mu_Bm\left(1+4\beta M\frac{e^2}{an^2}\right)
\end{equation}
where $\mu_B=e\hbar/(2Mc)$ is the Bohr magneton. We note that
correction to the magnetic moment depends only on one parameter of
deformation in the presence of two-parametric deformed algebra.

Let us compare our results with the corrections that follows from
the relativistic theory. The expression for the operator of an
orbital magnetic moment in this case reads \cite{davydov}
\begin{equation}\label{moment_relativ}
\hat{\mu}^{(rel)}=-\frac{e}{2Mc}\frac{Mc^2}{E_p}\hat{L}
\end{equation}
where $\hat{L}$ is the angular momentum operator and
$E_p=\sqrt{M^2c^4+\hat{p}^2 c^2}$ is the energy of a relativistic
particle. In the weak-relativistic approximation we can decompose
the energy of the particle in the series over $1/c^2$. Taking into
account only the first order terms we obtain
\begin{equation}\label{mag_moment_rel_approx}
\hat{\mu}^{(rel)}=-\frac{e}{2Mc}\left(1-\frac{1}{Mc^2}\hat{K}\right)\hat{L},
\end{equation}
where $\hat{K}=\hat{p}^2/2M$ is the kinetic energy operator of
non-relativistic particle.

The average value of the magnetic moment in a weak-relativistic
approximation is
\begin{equation}\label{magn_moment_rel2}
\mu^{(rel)}_z=\langle
nlm|\hat{\mu}^{(rel)}_z|nlm\rangle=-\frac{e}{2Mc}\left(1-\frac{1}{Mc^2}\langle\hat{K}\rangle\right)L_z
\end{equation}

Comparing relations (\ref{magn_moment2}) and
(\ref{magn_moment_rel2}) we see that corrections to the magnetic
moment take the similar form in both cases but contrary to
deformation the relativity theory gives the corrections with the
opposite sign. So the deformation of commutation relations leads
to the increase of the magnetic moment. At the same time the
relativity theory leads to the decrease of the magnetic moment of
the electron.

We rewrite relation (\ref{magn_moment3}) using two parameters
$\Delta x_{min}=\hbar\sqrt{\beta+\beta'}$ and
$\eta=\beta/(\beta+\beta')$ instead of $\beta$ and $\beta'$
similarly to \cite{Benczik,mykola,mykola2,mykola3}
\begin{equation}
\mu_z=\mu^{(0)}_z\left(1+\varsigma(\Delta x_{min},\eta,
n)\right),\quad \mu^{(0)}_z=-\mu_Bm
\end{equation}
where
\begin{equation}\label{varsigma}
\varsigma(\Delta x_{min},\eta,
n)=\frac{\mu_z-\mu^{(0)}_z}{\mu^{(0)}_z}=\eta\frac{4\Delta
x^2_{min}}{a^2n^2}
\end{equation}
is the specially introduced function which shows the corrections
to the orbital magnetic moment caused by deformation.

Using relation (\ref{varsigma}) we can numerically estimate
corrections to the magnetic moment caused by the minimal length
effects. Relation (\ref{varsigma}) shows that function $\varsigma$
depends on $\Delta x_{min}$, $\eta$ and $n$. As was shown in
\cite{mykola} the parameter $\eta$ has a bounded domain of
variation: $1/3\leqslant\eta\leqslant 1$. For the minimal length
we use the estimation obtained in \cite{Benczik,mykola,mykola2}
from the analysis of Lamb shift. Here we take $\Delta
x_{min}=10^{-16}$m.

\begin{figure}
\centerline{\includegraphics[scale=1,clip]{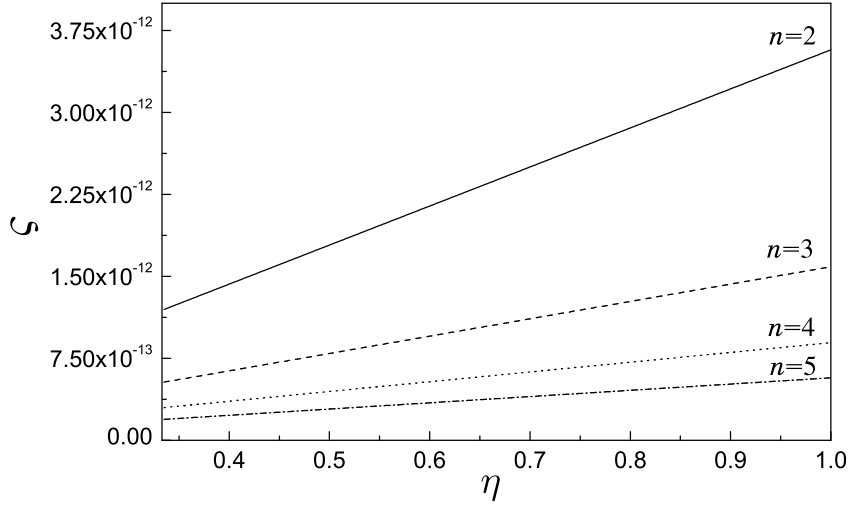}}
\caption{Correction to the orbital magnetic moment of the electron
in the hydrogen atom $\varsigma$ as a function of the parameter
$\eta$ for fixed values of the principal quantum number
$n$}\label{fig1}
\end{figure}

Figure \ref{fig1} shows a dependence of $\varsigma$ as a function
of $\eta$ on its domain of variation when the quantum number $n$
is fixed. We see that correction to the orbital magnetic moment
increases with the increasing of $\eta$ and decreases for higher
excited states. We compare the function $\varsigma$ introduced for
the estimation of minimal length effects with the relative error
of measurements of the Bohr magneton. The most recent measurements
show that the relative error for the Bohr magneton equals
$\varepsilon=2.5\times 10^{-8}$ \cite{CODATA}. Our estimation
shows that the upper bound for correction to the orbital magnetic
moment equals $\varsigma_{max}=3.57\times 10^{-12}$ that is four
powers less than the relative error of the corresponding error. So
we can conclude that Lamb shift at this moment gives the most
precise estimation of minimal length.

\section{Discussion}

We investigated the orbital magnetic moment of the electron in the
hydrogen atom in the deformed space with minimal length. Having an
explicit expression for the probability density flux we obtained
the electric current density for the electron in the hydrogen atom
and calculated corrections to orbital magnetic moment. The orbital
magnetic moment of the electron in the hydrogen atom depends only
on one deformation parameter in the presence of two-parametric
deformation. We showed that orbital magnetic moment in deformed
space is proportional to the angular momentum as in an ordinary
case. But the factor between the magnetic moment and angular
momentum is not constant as in an ordinary quantum mechanics. This
factor depends on the deformation parameter $\beta$ and the mean
value of kinetic energy of the electron and increases with the
increasing the kinetic energy. The kinetic energy of the electron
in the hydrogen atom is inversely proportional to the square of
the principal quantum number so the correction to the orbital
magnetic moment drops with the increasing of the principal quantum
number.

It is interestingly to note that correction to the magnetic moment
caused by deformation is similar to the corrections that follow
from the relativistic theory. But in contrast to the relativistic
theory deformation leads to an opposite sign of the correction.
Hence the relativity theory gives a negative correction to the
magnetic moment at the same time the deformation of commutation
relations leads to a positive one.

In order to estimate the correction to the orbital magnetic moment
of the electron caused by deformation we used the upper bound for
the minimal length obtained in \cite{Benczik,mykola,mykola2} from
the analysis of Lamb shift. We found that the upper bound for a
relative correction to the orbital magnetic moment
$\varsigma=(\mu_z-\mu^{(0)}_z)/\mu^{(0)}_z$ is $\sim 10^{-12}$. It
is four powers lesser than the relative error for the most recent
experimental value of Bohr magneton. So we can conclude that at
this time the measurements of the Bohr magneton are not enough
precise to obtain a more exact upper bound for the minimal length
in comparison with the minimal length that follows from the Lamb
shift.

\end{document}